# *KRAS* mutation testing in colorectal cancer as an example of the pathologist's role in personalized targeted therapy: a practical approach

Paweł Domagała[1], Jolanta Hybiak[1], Violetta Sulżyc-Bielicka[2], Cezary Cybulski[3], Janusz Ryś[4], Wenancjusz Domagała[1]

[1]Department of Pathology, Pomeranian Medical University, Szczecin, Poland
[2]Department of Clinical Oncology, Pomeranian Medical University, Szczecin, Poland
[3]Department of Genetics and Pathology, Pomeranian Medical University, Szczecin, Poland
[4]Department of Tumour Pathology, Centre of Oncology – Maria Skłodowska-Curie Memorial Institute, Kraków Branch, Poland

Identifying targets for personalized targeted therapy is the pathologist's domain and a treasure. For decades, pathologists have had to learn, understand, adopt and implement many new laboratory techniques as they arrived on the scene. Pathologists successfully integrate the results of those tests into final pathology reports that were, and still are, the basis of clinical therapeutic decisions. The molecular methods are different but no more difficult to comprehend in the era of "kit procedures". In recent years, the development of targeted therapies has influenced routine practices in pathology laboratories because the use of molecular techniques is required to include clinically useful predictive information in the pathology report. Pathologists have the knowledge and expertise to identify particular gene mutations using the appropriate molecular tests currently available.

This review focuses on the most important recent developments in *KRAS* mutation testing in metastatic colorectal cancer (CRC), and shows that a pathologist is involved in 10 stages of this procedure. Recent studies have shown that highly sensitive, simple, reliable and rapid assays may significantly improve the identification of CRC patients resistant to anti-EGFR therapy. Thus, direct sequencing does not seem to be an optimal procedure of *KRAS* testing for clinical purposes. Twelve currently available high-sensitivity diagnostic assays (with the CE-IVD mark) for *KRAS* mutation testing are briefly described and compared. The suggested pathology report content for somatic mutation tests is described. In conclusion, evidence is presented that sending away paraffin blocks with tumor tissue for *KRAS* mutation testing may not be in the best interest of patients. Instead, an evidence-based approach indicates that *KRAS* mutation testing should be performed in pathology departments, only with the use of CE-IVD/FDA-approved *KRAS* tests, and with the obligatory, periodic participation in the *KRAS* EQA scheme organized by the European Society of Pathology as an independent international body.

Introduction

For decades, pathologists have had to successfully learn, understand, adopt and implement many laboratory techniques as they arrived on the scene (e.g., histological, cytological, histochemical, ultrastructural and immunohistochemical techniques). Pathologists were able to integrate the results of those tests into final pathology reports that were, and are, the basis of clinical therapeutic decisions. This does not mean that pathologists had to perform those laboratory procedures themselves; rather, they had to understand them and know how and when to apply them. But, who performed the tests? Biologists employed in pathology laboratories performed the tests. We do not see the basic difference with the emergence of new molecular methods. Again, with the advent of molecular methods pathologists have to understand them and know how and when to apply them. In many pathology laboratories this understanding has already occurred. In fact, a continuing education is the first commandment for a physician in general and the pathologist in particular. Such education may be painful for some, but it is necessary. Of course, the integration of new methodology has always been a team endeavor; specifically, biologists performed the procedures and pathologists integrated the results to formulate diagnoses. In our view, molecular methods are different but they are not more difficult to comprehend than previous methods. Molecular methods can be understood, adopted and implemented by an average pathologist who can easily learn how to integrate the results into a pathology report, provided that appropriate training programs in genomic medicine are organized during the pathology residency. Molecular biology is not a hermetic, mysterious knowledge that one cannot master. In fact, clinically useful molecular biology techniques are not especially difficult to learn because nowadays many molecular procedures have been standardized and greatly simplified. To obtain proper results, just one major requirement has to be fulfilled, namely faithful adherence to the detailed manufacturer's instructions. It is obvious that pathologists have to cooperate with the molecular biologists employed in their departments. However, in our view, it is equally important that pathologists learn the basics of molecular biology necessary to implement molecular biology procedures in their laboratories. Therefore, continuous efforts should be made in this respect during the pathology residency period, as well as afterward, in the process of continuing education.

Pathological diagnosis

In recent years, the development of targeted therapies has influenced the routine practices used in surgical pathology laboratories. Previously, cancer therapy was based on the histological type and grade of a tumor as well as the clinical stage of the disease. With the advent of targeted therapy, specific predictive information is sought. The results of an assessment of several predictive markers (mutations, amplifications, and proteins) have to be incorporated into the traditional pathological diagnosis because not all patients derive clinical benefit from targeted therapy. A classic example of targeted therapies is anti-estrogen therapy, which is successful only in patients with breast cancer cells expressing estrogen/progesterone receptors. Thus, a pathologist must provide a specific diagnosis and is expected to assess predictive markers so that patients most likely to respond to targeted therapies may be identified. Furthermore, when several protein/molecular targets are identified in a particular tumor, the development of personalized therapy (in the strict sense of the word) is possible because several therapeutic options may be revealed that are available only for a given patient [1].

**Take-home messages:**

- *KRAS* mutation status has emerged as an important predictive marker for anti-EGFR therapy in patients with mCRC
- A pathologist is involved in *KRAS* testing in 10 stages of the procedure
- Microdissection should be performed in every case to obtain tissue enriched with tumor cells for *KRAS* mutation testing
- The sensitivity of direct sequencing for analysis of *KRAS* mutations in FFPE tumor samples seems to be too low to enable its use as a routine clinical test
- The final choice of the CE-IVD-marked *KRAS* mutation test is largely dependent on the laboratory equipment, experience and cost of the test
- *KRAS* mutation testing should be performed in pathology departments, only with CE-IVD/FDA-approved *KRAS* tests, and with obligatory, periodic participation in the *KRAS* EQA scheme organized by the European Society of Pathology as an independent international body

Currently, predictive immunohistochemical and molecular diagnostic tests identify suitable patients for targeted therapy. Therefore, to include predictive information in the pathology report, immunohistochemistry and molecular techniques are required. This simply means that to provide oncologists with meaningful predictive information (incorporated into the pathological diagnosis), contemporary pathology departments should be equipped with appropriate laboratories where molecular techniques necessary to reveal targets for implementation of targeted therapy can be performed, similar to the setup of contemporary immunohistochemical laboratories. Molecular tests can also be performed outside the pathology department in certified molecular laboratories. However, in such situations:

1) there are unnecessary delays associated with the need to send sections or paraffin blocks, which is not in the best interest of patients;
2) molecular biologists who perform a test in a molecular laboratory may be unfamiliar with the latest clinical data related to the molecular alterations tested and the intricacies of pathological, clinical and therapeutic significance of particular tests;
3) responsibility for the final diagnosis is blurred;
4) the pathologist has a curious role of a distributor of paraffin blocks/sections and a messenger of predictive information between a molecular biologist and a physician. At best, this role is that of a coordinator rather than of a generator of the vital final pathological diagnosis with predictive information.

It may be helpful here to define the meaning of the term "pathological diagnosis" of tumors. **A pathological diagnosis (report) is obtained on the basis of microscopic analysis of tissue and/or cells TOGETHER WITH additional tests (e.g., histochemical, immunohistochemical and, molecular tests) and clinical data, to provide information concerning histological type of a tumor AND predictive/prognostic data.**

## *KRAS* mutations

There are three *RAS* genes in the human genome; *KRAS*, *HRAS* and *NRAS*. Approximately 15-20% of all human neoplasms contain *RAS* mutations. Mutations in the *KRAS* gene are detected in 35-45% of CRCs [2-8], whereas *NRAS* and *HRAS* mutations are only found in 1-3% of CRCs. *KRAS* mutations are frequent in pancreatic, colorectal, biliary tract, and lung cancers. The most common *KRAS* mutations in CRCs are found in codons 12 (~77% of mutations) and 13 (~20% of mutations) in exon 2 of the gene. Much less frequent are mutations in codons 61 (1%) and 146 (0.5%) (data from the COSMIC database v60) [9]. Somatic missense mutations in codon 12 of the *KRAS* gene, resulting in a single amino acid substitution (p.Gly12Val), are the most common abnormalities in CRCs.

## RAS/EGFR signaling pathway

The *KRAS* gene encodes a small GTPase protein (G protein) that functions downstream of EGFR-induced cell signaling, and participates in the activation of important oncogenic signaling pathways. The RAS protein is a key element of the RAS-MAPK intracellular signaling pathway, which is triggered by EGFR tyrosine phosphorylation of the intracellular domain of the receptor. This triggering results in the activation of several other signaling pathways that control gene transcription, cell proliferation, apoptosis, angiogenesis, invasion and migration [10]. Therefore, the EGFR signaling system is regarded as crucial to the regulation of the growth, proliferation and malignant transformation of epithelial cells.

RAS proteins, which act as signal transducers, normally cycle between active (RAS-GTP) and inactive (RAS-GDP) conformations. RAS proteins are activated by guanine nucleotide exchange factors (GEFs) and inactivated when RAS-GTP is hydrolyzed to RAS-GDP by GTPase-activating proteins (GAPs). In normal cells, the activities of GEFs and GAPs are tightly controlled and the RAS-GTP level is kept in check. Mutations in *RAS* result in markedly reduced the intrinsic GTPase activity of RAS proteins and make these proteins resistant to the GAPs. Thus, mutated RAS proteins permanently remain in the RAS-GTP active form and continuously activate signaling pathways without stimulation of the HER family of cell surface receptors [11-14]. Such an activating mutation may induce an oncogenic transformation or confer resistance (insensitivity) to anti-EGFR antibody therapies due to permanent, EGFR-independent activation of both the PI3K/AKT and MAPK pathways.

## Mutations in *KRAS* as predictive markers for anti-EGFR treatment in patients with metastatic colorectal cancer

Given the EGFR independent activation of *KRAS* it is not surprising that, anti-EGFR antibodies will not be effective in patients with metastatic CRC (mCRC) with a *KRAS* mutation. Indeed, an association of *KRAS* mutations in CRC with resistance to anti-EGFR monoclonal therapy was reported [15, 16]. These results were subsequently confirmed in independent retrospective studies [17, 18]. Finally, supportive evidence of predictive significance of *KRAS* mutations for anti-EGFR treatment of mCRC was obtained from phase II and III clinical trials of the anti-EGFR monoclonal antibodies cetuximab and panitumumab used either as monotherapy or in combination with chemotherapy in patients with mCRC (references in [19]). For ex-ample, single-agent, randomized, controlled trials with cetuximab [20] and panitumumab [21] revealed no benefit from these anti-EGFR therapies in *KRAS*-mutant mCRCs patients. In a study of second- or third-line treatment of mCRC patients with cetuximab, the statistically significant differences between the antibody arm and control arm were 12.8% vs. 0% for response rate, 3.7 months vs. 1.9 months for progression-free survival and 9.5 months vs. 4.8 months for median overall survival [20]. In April 2009, based on an analysis of five randomized controlled trials of cetuximab or panitumumab and five single-arm retrospective studies, the following provisional clinical opinion of the American Society of Clinical Oncology was issued: "…all patients with mCRC who are candidates for anti-EGFR antibody therapy should have their tumor tested for *KRAS* mutations in a CLIA-accredited laboratory. If a *KRAS* mutation in codon 12 or 13 is detected, patients with mCRC should not receive anti-EGFR antibody therapy as part of their treatment" [19]. Both the European Medicines Agency (EMEA) and the U.S. Food and Drug Administration (FDA) approved the use of cetuximab and panitumumab for treatment in patients with wild-type *KRAS* mCRC. However, there are differences in those approvals. The EMEA approved both drugs for the treatment of patients with wild-type *KRAS* mCRC in 2007 without mentioning methodology or the exact mutations to be tested. In contrast, the FDA specified the testing for mutations in codon 12 or 13 of *KRAS* in their recommendations issued in 2009 and, in addition, required that two conditions be met: validation of a single assay for mutation detection and reassessment of all randomized clinical trials with this assay. In July 2012, the FDA approved the TheraScreen kit to be sold as a companion diagnostic test for cetuximab. Furthermore, the FDA has additionally approved the use of cetuximab in combination with FOLFIRI (5-fluorouracil, irinotecan, leucovorin) as a first-line treatment in patients with mCRC and EGFR-expressing and wild-type *KRAS* tumors. Thus, **KRAS mutation status has emerged as an important predictive marker for anti-EGFR antibody therapy in patients with mCRC.** The clinical trial data briefly summarized above strongly support the recommendation that patients whose CRCs do not have *KRAS* mutations should be treated with anti-EGFR monoclonal antibodies, and that those with *KRAS* mutations should not be treated with this therapy. The results described above pertain to *KRAS* mutations in codons 12 and 13 of exon 2. Recent reports suggest that although rare *KRAS* exon 3 mutations may also be associated with a lack of response to anti-EGFR monoclonal antibodies [22]. However, the examination of *KRAS* mutations in exon 3 does not significantly improve the identification of non-responding mCRC patients.

Thus, the presence of *KRAS* mutations has become widely accepted as a negative predictive marker for anti-EGFR monoclonal antibody therapy in mCRC patients. However, in the future, the presence of these mutations may also become a positive predictive marker for therapies based on the inhibition of *KRAS* activation or the inhibition of alternative downstream kinases in *KRAS*-mutant mCRC patients. There are several ongoing preclinical studies and clinical trials utilizing different therapeutic approaches in this respect, e.g., BRAF inhibitors in *BRAF*-mutant mCRCs, MEK inhibitors, farnesyltransferase inhibitors, and SFK inhibitors (e.g., dasatinib), which may sensitize *KRAS*-mutant mCRCs to cetuximab (dual kinase inhibition) [23]. These new therapeutic approaches only underscore the clinical importance of *KRAS* testing for the targeted therapy of patients with mCRCs.

### Key role of the pathologist in the assessment of therapeutic targets in general and *KRAS* mutation testing in particular

A pathologist is involved in *KRAS* testing in 10 stages of the procedure.

1) *A pathologist decides on the method of tumor tissue collection for KRAS testing prior to DNA extraction.* There is no consensus recommendation for this step, although general recommendations have been made (by CAP [24] and NCCN [25]) regarding the process of tissue collection. From a practical point of view, buffered formalin-fixed, paraffin-embedded (FFPE) tumor tissue is most commonly used for *KRAS* testing. Testing may also be performed on fresh tissue stored in a preservative solution (e.g., RNAlater {Ambion}), rapidly frozen and stored tissue [19] or formalin-free fixed tissue [26, 27].

2) *A pathologist identifies sources of tumor material for the assessment of KRAS mutation status.* According to the National Comprehensive Cancer Network's (NCCN) clinical practice guidelines [25], either primary or metastatic tumors can be used for *KRAS* testing. Based on results showing a high level of concordance between *KRAS* mutation status in primary vs. metastatic CRCs [28, 29] and observations indicating that *KRAS* mutations may occur early in the progression of CRC, there is a consensus opinion that in metastatic disease *KRAS* testing of the primary CRC provides conclusive results. Because archival paraffin blocks from the primary tumor are available in the majority of cases of metastatic disease, there is no need for biopsy of a metastatic tumor. When a patient presents with metastasis at the initial diagnosis, a metastatic tumor biopsy may be used for the *KRAS* test. When post-radiochemotherapy low tumor cellularity is found in a specimen, pre-treatment biopsies repre-

3) *A pathologist is responsible for the selection and evaluation of the tumor tissue block.* A pathologist makes the judgment as to whether there is sufficient quantity and quality of tumor material to be used in DNA extraction for *KRAS* testing (see below). The amount of tumor tissue in the specimen and the general quality of the collected tissue and extracted DNA have important implications for the accuracy and sensitivity of the *KRAS* testing method used [31].

4) *A pathologist is responsible for tumor cell enrichment in a target tissue sample.* To increase the sensitivity of mutation testing, several methods may be employed to enrich the tested sample in tumor cells, e.g., macrodissection, manual microdissection or laser microdissection (see below). All of these techniques require the user to be familiar with microscopy and histopathology [32].

5) *A pathologist chooses the method to be used for KRAS testing in the laboratory.* To make a well-informed decision regarding the testing method to be used, the pathologist should consult with a molecular biologist and take into consideration the DNA amount and quality required, time to obtain results, accuracy, sensitivity and the cost of the preferred method (see below and Table I). Each method has its advantages and disadvantages. However, although traditional Sanger sequencing detects all clinically-important *KRAS* mutations, its sensitivity is too low to be used as a routine clinical test (mutant alleles must be present in at least 30-40% of cells for reproducible detection) [2, 33-35]. To the best of our knowledge, there are currently 12 highly sensitive, simple, rapid CE-IVD-marked diagnostic assays for *KRAS* mutations (see below and Table I). The minimal requirements that a *KRAS* test should satisfy include: a specificity of 100% and a mutation detection sensitivity of between 95 and 99% [36]. It is clear that increasing the sensitivity of the methods used to detect *KRAS* mutations can greatly improve predictions of resistance to anti-EGFR treatment [37-41]. Therefore, **high sensitivity of the method is a key issue.** Due to the intratumoral heterogeneity of *KRAS* mutations [42-45], clones bearing *KRAS* mutations may be undetected by direct sequencing [37, 41], and tumor cells from these clones may exhibit an increased propensity for distant metastases [37].

6) *A pathologist, together with a molecular biologist, actively participates in the laboratory procedure for KRAS testing.* Pathologists should be aware of the types of procedures used and their pros and cons, and specifically, they should be familiar with the details of the procedures used in his or her department. Test data should be analyzed and discussed

**Table I.** Comparison of *KRAS* genotypes detected by CE-IVD-marked commercial kits, frequency of their appearance in CRCs and cost per sample. Green boxes highlight mutations that are detected by a particular kit

| CE-IVD *KRAS* KIT | APPROX. COST PER SAMPLE (USD)[a] | A.A. Mutations and their frequency in CRC (%)[b] | | | | | | | |
|---|---|---|---|---|---|---|---|---|---|
| | | G12A, G12C G12D, G12R G12S, G12V G13D (96.2) | G13C (0.5) | G13S G13R (0.3) | G13A G13V G13I G12F (0.3) | Q61H (CAT) Q61H (CAC) (0.5) | Q61L Q61R (0.3) | Q61K Q61E Q61P (0.1) | Other |
| TheraScreen PCR | 210 | ✓ | | | | | | | |
| AmoyDx | 117 | ✓ | | | | | | | |
| PNAClamp | 70 | ✓ | | | | | | | |
| RealQuality | 50 | ✓ | | | | | | | |
| EntroGen | 77 | ✓ | | | | ✓ | ✓ | | |
| LightMix | 57 | ✓ | ✓ | ✓ | | | | | [d] |
| StripAssay | 77 | ✓ | | ✓ | | | | | [e] |
| Hybcell plexA[c] | 124 | ✓ | | | | | | | |
| Devyser[c] | 40 | ✓ | | | ✓ | ✓ | | | |
| Surveyor[c] | 26 | ✓ | | | | | | | [f] |
| Cobas[c] | 81 | ✓ | | | | ✓ | ✓ | ✓ | |
| TheraScreen Pyro[c] | 124 | ✓ | | | | | | | [g] |

[a]According to price quotes (net) available in Poland (August 2012). Cost calculated per sample when the maximum number of cases is tested in a batch (the optimum circumstances). This is only the kit cost per sample and does not include the costs of shipping, DNA isolation and quantification, repeat tests or labor cost; [b]The mutation frequency (CRCs only) was obtained from the COSMIC database v60; [c]Requires equipment other than that used for real-time PCR; [d]G12T; [e]G12I and G12L; [f]other mutations in exon 2; [g]other mutations in codons 12, 13, and 61

149

Table II. Pathology report content for a somatic mutation test. A checklist

| General information |
|---|
| ❑ Report title[1] |
| ❑ Name and address of the reporting institution |
| ❑ Name and address of the requesting institution and name of the requesting oncologist |
| ❑ Date of the request form (if material comes from the same pathology department where mutation test will be performed) or date of arrival of the sample with the request form |
| ❑ Date of the report |
| ❑ Own reference number of mutation test |
| ❑ Total number of pages and page numbers (e.g., 1 of 2) *CPT codes[2]* *Cite references for report facts* |

| Patient information |
|---|
| ❑ Name and age |
| ❑ ID number or date of birth *Short clinical history* |

| Sample information |
|---|
| ❑ Sample number |
| ❑ Nature of the sample (e.g., FFPE[3], fresh frozen, or biopsy) |
| ❑ Tumor origin (e.g., primary, metastatic, or recurrent) |
| ❑ Sample adequacy, e.g., satisfactory for evaluation or unsatisfactory for evaluation (specify reason) *Percentage of tumor cells[4]* |

| Methods used |
|---|
| ❑ Identification of the best block with the highest percentage of tumor cells: yes or no |
| ❑ Method of tumor cell enrichment: macrodissection, microdissection, or none |
| ❑ Method of DNA isolation |
| ❑ Method of genotyping |
| ➢ Name of the diagnostic kit (CE-IVD/FDA-approved) |
| ➢ Sensitivity of the method |
| ➢ List of mutations that were tested[5] |

| Result |
|---|
| ❑ Genotype[5] |
| ❑ Interpretation regarding the treatment[6] |
| ❑ Comments[7] (if applicable) |

| Responsibility |
|---|
| ❑ Signatures[8] and printed names of the molecular biologist and pathologist involved in performing the test and interpreting the results |

*The optional items in each section are shown (without a checkbox) in italics.*
[1]*Should include the name of the tested gene(s) clearly distinguished from the rest of the report;* [2]*Current Procedural Terminology (CPT) codes document the analytical and interpretative procedures that are performed in the laboratory;* [3]*Formalin buffered or not;* [4]*If the test was performed without microdissection and/or a low sensitivity method was used;* [5]*Nomenclature according to HGVS guidelines;* [6]*Example: presence of KRAS mutation indicates likely resistance to anti-EGFR therapy;* [7]*Examples: recommendations for further testing; condition of a specimen that may limit the adequacy of testing; reason why a specimen was rejected or not processed to completion; if the report is an amended or addendum report, a description of the changes or updates;* [8]*Reports may be signed electronically*

jointly by the pathologist and molecular biologist. The pathologist is also involved in the continuous review process of new technologies and publications on *KRAS* mutation testing.

7) ***A pathologist writes the results of a KRAS test in a special report that is understandable to the oncologists.*** The *KRAS* mutation report, similar to the report of any molecular test of clinical significance, should contain items listed in Table II [19, 36, 46-54]. If a mutation is identified, the affected codon and the specific change should be reported. This is an important aspect of the report because recommendations for treatment eligibility are continuously evolving.

8) ***A pathologist provides the interpretation of results in terms of their predictive significance.*** The pathologist should report the results of *KRAS* testing, as well as their therapeutic significance, in the particular clinical context of a patient. The interpretation of results should be clear and accurate in view of the findings described above (in the paragraph "Mutations in *KRAS* as predictive markers for anti-EGFR treatment in patients with metastatic colorectal cancer") and the most recent literature data. The interpretation of the genomic information should assist a physician in therapeutic decision-making, i.e., to help predict which patients are most likely to benefit from anti-EGFR monoclonal antibody treatment. Specifically, it should be noted that a wild-type result does not rule out the presence of rare mutations not tested by the assay used. Moreover, it should be added that not all patients with wild-type *KRAS* cancers will respond to anti-EGFR antibody therapies, the predictive information of the *KRAS* test applies only to anti-EGFR monoclonal antibody therapy and not to other types of therapies, and *KRAS* mutation tests in metastatic and corresponding primary cancers are discordant in less than 5% of cases [55].

9) ***A pathologist is responsible for the turnaround time of the report.*** A recent survey of *KRAS* testing revealed that test results were available within 15 days for 82%, 51% and 98% of tested patients in Europe, Latin America and Asia, respectively [56]. In another study, the results were obtained with a mean delay of 33 days [57]. Although it seems that a delay of 10-14 working days is acceptable [47], an effort should be made to make the report available in a shorter time (one has to keep in mind that a patient with metastatic CRC is waiting for this test and the therapeutic decision). **The turnaround time can be substantially shortened if the whole procedure is performed in the pathology department that provided the histopathological diagnosis of CRC.** In fact, in our experience, *KRAS* mutation test can be done and the report issued **in three working days**, starting from the date on which the test was requested.

10) *A pathologist is responsible for the implementation of an external quality assurance (EQA) program in the molecular diagnostic laboratory of the pathology department in which he or she works.* We share the well-founded opinion that evidence-based approaches require EQA programs for *KRAS* testing (see below).

In summary, the pathologists looking down the microscope remain the final decision makers [58]. **They may seek advice from molecular biologists and oncologists but the final diagnostic decision and the responsibility is theirs.** Certainly, "…genomic analysis in the absence of pathologic assessment risks the generation of meaningless results" [59]. Currently a pathology report that is meaningful to the oncologist as well as to the patient includes a diagnostic algorithm integrating conventional histopathology with immunohistochemistry [60] and disease-specific molecular tests.

## *KRAS* testing as a routine test necessary to provide predictive information in pathology reports of colorectal cancer

Because the assessment of *KRAS* mutations has become an important aspect of management of mCRC patients, there is an urgent need to establish and agree on the best laboratory procedures to ensure accurate assessment of *KRAS* status. A routine predictive test should be relatively simple to perform, fast, reliable (in terms of sensitivity and specificity) and not too expensive. **Contemporary somatic mutation testing has taken the form of a "kit procedure".** It is simple, fast and reliable, provided that a test that is approved by the appropriate national or international agencies is used, the producer's protocol is rigorously followed, and an EQA program is implemented by the laboratory. Below, we briefly describe the most important aspects of *KRAS* mutation testing procedures in which FFPE CRC tissue is used. Currently, FFPE tissue is widely used in *KRAS* testing, but fresh-frozen tissue or cytological material can also be clinically useful.

### Selection and evaluation of the tumor tissue block

Although somatic mutation analysis does not require a pure cell population, tumor cellularity is a critical issue. A pathologist responsible for selecting a paraffin block for molecular tests should choose one that has the greatest percentage of invasive cancer cells and avoid blocks with many lymphocytes, necrosis or extracellular mucin. These factors can diminish the sensitivity of the technique, particularly if direct sequencing is used [61]. It has to be kept in mind that genotyping is performed on genomic DNA of both tumor cells and benign stromal cells of tumor microenvironment. Thus, sections for the test should be taken from blocks containing tumor tissue enriched with intact cancer cell nuclei. International guidelines suggest that the specimen should contain a cancer cell percentage of at least 70% if a low-sensitivity technique, such as direct sequencing of the PCR product, is used [46]. In any case, the specimen should contain at least 100 cancer cells [62].

Evaluation of the tumor's contents is a subjective procedure that depends on both the region selected and the assessment of the ratio between the cancer cells and normal nucleated cells. High inter-observer variability has been reported in the estimation of the percentage of cancer cells in hematoxylin and eosin (HE) stained sections. An EQA study involving 13 experienced laboratories revealed large differences in the estimates of this percentage in the same samples, varying from 9 to 90% or from 32 to 95%. There was no consistent pattern in the estimates between different laboratories [31, 47, 63]. These differences may be partially explained by incorrect measurement of the percentage of the sample area that is tumor, instead of the percentage of nuclei that are in the tumor [64], which is more difficult to measure with the naked eye. Furthermore, large differences in the diameter of tumor cell nuclei and the nuclei of the neighboring benign stromal cells make the estimation of tumor DNA content difficult and imprecise, and usually the contribution of normal cell nuclei to total DNA content is underestimated. Thus, the histological assessment of tumor cell percentage can serve only as a very rough estimate of tumor DNA content in the corresponding extract [65, 66]. For the above reasons, tumor cell enrichment is necessary for genotyping with FFPE tissue.

### Macrodissection and microdissection

The percentage of cancer cells in the specimen and the general quality of the collected tissue and extracted DNA have important implications for the accuracy and sensitivity of the *KRAS* testing method used [31]. The target tissue will always contain some normal cells within the tumor microenvironment (e.g., lymphocytes, macrophages, endothelial cells, and fibroblasts). To increase the sensitivity of mutation testing, several methods may be employed to achieve cancer cell enrichment in tested material, e.g., macrodissection, laser microdissection, and manual microdissection.

**Macrodissection.** In this method, HE-stained slides are examined and the enriched tumor cell area is marked by a pathologist on a HE-stained slide that will be used as a template for macroscopic dissection (without the microscope) of the consecutive slide(s). The area corresponding to the area marked on the template HE slide is scratched off using a sterile, single-use scalpel.

**Laser microdissection.** This is the most accurate technique to separate tumor cells from benign stromal cells. However, it is time-consuming and requires very expensive equipment. Therefore, it cannot be recommended for routine use in every case. In our experience and that of others [67], in some cases where tumor cellularity is low or tumor cells are dispersed (e.g., small biopsy, cytology specimens, or post-radiochemotherapy samples), isolation by laser microdissection results in a purer population of tumor cells for mutation detection and is a more sensitive method compared with other methods.

**Manual microdissection.** This simple technique is more accurate than macrodissection and less meticulous than laser microdissection. It can be performed on unstained tissue, but the best results are obtained using stained tissue sections, enabling the procurement of tissue compartments (even as small as 1 mm$^2$) much faster than laser microdissection and more accurately than macrodissection [68]. For manual microdissection, the tumor area is dissected by a pathologist under a light microscope (standard or inverted). The compartments without cancer cells are removed with a scalpel or injection needle, and the remaining tissue fragments (containing nearly all cancer cells) are removed (scratched) into a test tube for further treatment (one can also place the cancer cells into a test tube first). Almost any histological staining (e.g., cresyl violet, methyl green, methylene blue, toluidine blue, and nuclear fast red) can be used to visualize nuclei in sections selected for DNA or RNA molecular analysis, because the staining does not interfere with DNA and RNA testing [69]. Although some older studies reported that HE staining inhibited DNA amplification by PCR [70, 71], recently published data [69, 72, 73] provide no evidence for the interference of HE staining with DNA testing. These results suggest that DNA from HE-stained sections can be effectively used for routine DNA testing, which is especially important when only HE-stained slides, and not tissue blocks, are available. Then, after the HE slides are scanned, somatic mutations can be directly tested from routinely stained HE sections.

Manual microdissection is a powerful tool to enrich the analyzed sample with tumor cells. Recent research on the reported clinical benefit associated with highly sensitive *KRAS* detection methods suggests that manual microdissection will become an indispensable routine method to obtain tissue enriched with tumor cells for the analysis of somatic mutations. In this approach, subjective counting of the relative frequency of neoplastic cells in a tissue sample may be avoided because manual microdissection, together with highly sensitive diagnostic *KRAS* tests, assures a high level of sensitivity and repeatability. Thus, in our opinion, **manual microdissection should be performed in every case to obtain tumor tissue of the appropriate quality and quantity for *KRAS* mutation testing.**

**DNA isolation**

DNA quality is one of the most important factors when performing DNA mutation assays in FFPE tissue; therefore, the DNA extraction methodology is critical. The formation of DNA-protein cross-links due to the formaldehyde solution leads to nucleic acid fragmentation but at the same time, nucleases are deactivated (- a stabilizing effect). Thus, while molecular templates from FFPE tissue are of inferior quality compared with those from fresh-frozen tissue, they may still be useful for nucleic acid assessment with recently developed methods (even for microarray profiling and wide genome scans), and results from FFPE tissue are largely comparable with data obtained from fresh-frozen material [65, 74-76]. The main advantage of using FFPE tissue for molecular analyses in cancer research is the possibility of achieving an accurate correlation of the results with tissue histology and with long-term follow-up data. This ability is especially important in rare subtypes or subpopulations of cancers. Another advantage is the availability of FFPE tumor tissue for the assessment of predictive markers when new targeted therapies are developed in the future.

There are plenty of "in-house" methods and commercially available kits specially designed for DNA isolation from FFPE tissue. The aim of these methods is to increase the DNA yield from FFPE tissue, and they involve steps to reverse modification of nucleic acids by formaldehyde. The highest amount of DNA was obtained using the phenol-chloroform extraction method [73, 77-79] or the WaxFree DNA kit (TrimGen, Sparks, MD, USA) [78, 79]. However, the quality of the real-time PCR reactions is quite often compromised when using isolates with increased quantities of DNA [78]. Furthermore, spectrophotometric analysis revealed that DNA extracted with the use of the WaxFree DNA kit was of poorer quality compared with that obtained with the use of phenol-chloroform, silica-based columns (QIAamp DNA FFPE Tissue kit, Qiagen, Hilden, Germany) or rapid glass-fiber filters (RecoverAll kit, Ambion, Austin, TX, USA) [79]. Munoz-Cadavid *et al.* [80] recently evaluated the extraction of high-quality DNA from FFPE tissue in 5 commercial kits and found that the best results were achieved with the TaKaRa Dexpat Kit (Takara Bio, Shiga, Japan) and the QIAamp DNA FFPE Tissue Kit. The Qiagen extraction kit was the most popular DNA extraction method for FFPE tissue in the European Society of Pathology (ESP) *KRAS* EQA program [47].

DNA obtained from FFPE samples that allow the amplification of 300 bp is considered to be good-quality DNA and can be used for many molecular tests [81]. For tests that require a standardized input of DNA (e.g., multiplex PCR, MLPA, and array-CGH), isolation methods based on silica columns should be recommended. For general molecular tests (e.g., mutations and translocations), a simple, rapid and less expensive protocol for DNA isolation (without precipitation or purification) may be sufficient [81-83].

### Assessment of DNA quality and quantity

Widely used methods based on measurements of UV absorbance do not always allow for an assessment of the availability of a material for molecular tests. Common biological contaminants of extracted DNA such as proteins, RNA, and chaotropic salts from extraction procedures can increase the spectrophotometric estimation of DNA concentration [84]. Imprecise measurements may be particularly harmful when further molecular analysis requires the precise assessment of DNA content in the material studied.

Therefore, for the assessment of the quality of DNA extracted from FFPE tissue that is to be used for molecular tests, methods based on DNA-binding dyes fluorescence, real-time PCR DNA amplification or PCR with agarose gel electrophoresis are more useful than spectrophotometric analysis. The main advantage of real-time PCR-based analysis is the ability to assess how much of the available nucleic acid sample is amplifiable. Drawbacks of this method include the expensive proprietary reagents, primers, and probes needed to perform the assays, the lengthy assay time, and the fact that sample volume has to be expanded [85].

Fluorometric measurement of DNA concentration has gained popularity because it is simple and much more sensitive than absorbance measurements [86]. Fluorometric analysis is not as accurate as real-time PCR, but it is much faster, less expensive and simpler; therefore, it appears to be the optimal approach to the quantification of DNA for molecular analysis of FFPE tissue.

### Commercially available real-time quantitative PCR kits for clinical KRAS mutation analysis of FFPE tissue as a better alternative to direct sequencing

Several molecular techniques characterized by different sensitivities, specificities and complexities are currently used in research and clinical studies for the detection of KRAS mutations in FFPE tumor samples. Some techniques require expensive equipment and reagents, whereas other methods can be easily adapted for use in molecular laboratories of departments of pathology without the need for additional machines [87, 88]. Until recently, the most widely available methods for KRAS testing were Sanger sequencing and a variety of "in-house" developed laboratory tests. However, these methods are subject to great inter- and intra-laboratory variability and are not always prone to adequate Quality Control schemes that ensure reproducibility of results [89].

In general, the advantage of sequencing-based methods is the ability to detect all clinically important KRAS mutations. However, direct sequencing has several limitations: (1) it is time consuming and labor intensive; (2) interpretation is subjective if the signal/noise ratio is low; and (3) it has low sensitivity compared with other methods. A mutant allele percentage of more than 5 to 10% is required for detection of mutant tumor cells in pyrosequencing, and a percentage of 30 to 40% is needed for Sanger sequencing [33, 34, 61, 90]. The detection of mutations by direct sequencing of cytological specimens, small tumor biopsies or tumors containing a high percentage of non-neoplastic cells, may lead to false-negative results [88]. Bando *et al.* [2] suggested, on the basis of their experience and literature data, that direct sequencing should be discarded as the method of choice in clinical KRAS testing because with this method, approximately 25% of patients selected for anti-EGFR therapy will not receive any benefit. Thus, **the sensitivity of direct sequencing seems to be too low to be used as a routine clinical test for the analysis of KRAS mutations in FFPE tumor samples.**

The development of commercial real-time quantitative PCR assays offers a useful alternative because they are simple, labor-saving (in our experience the procedure takes about 3 h) and more sensitive than direct sequencing [2, 37-41, 61, 88, 91-100]. Furthermore, because they avoid post-PCR handling, the risks of contamination and generation of false-positive results are greatly reduced [101]. Real-time PCR methods yield informative results even in cases with very fragmented DNA, whereas only samples with relatively well-preserved DNA can be accurately analyzed with direct sequencing [65]. The per-sample cost of mutation test with clinically validated methods (CE-IVD-marked) is significantly higher compared with the per-sample cost using direct sequencing. However, costs of labor, assay development and validation of the method are not included in the assessments of the sequencing method. Moreover, basic equipment for real-time quantitative PCR tests is less expensive (than that for sequencing), and the molecular tests can be performed in most pathology departments with molecular diagnostic laboratories with real-time PCR machines.

### Clinical significance of high-sensitivity KRAS testing

In some CRCs, KRAS mutations may be present in only a small subset of tumor cells (clonal heterogeneity). The failure to find such mutated clones that may be associated with treatment resistance, substantially limits the ability to predict treatment response [41, 91, 102].

Recent studies have shown that highly sensitive, simple, reliable and rapid assays may significantly improve the identification of CRC patients resistant to anti-EGFR therapy. With such techniques, it is possible to detect *KRAS* mutations in a subset of tumors that seem to be wild-type when tested with direct sequencing. Using high-sensitivity methods, Molinari *et al.* [37] identified up to 11% additional *KRAS* mutations compared with direct sequencing, and Malapelle *et al.* identified 8% more *KRAS* mutations in patients not responding to cetuximab. Similarly, it has been reported that *KRAS* mutation testing by highly-sensitive methods and quality-controlled *KRAS* assays, may more closely correlate with the clinical effects of anti-EGFR antibody therapy than direct sequencing [38, 39], and it has also been reported that such mutation testing is useful for identifying true responders to cetuximab [40]. It should be noted that anti-EGFR monoclonal antibodies have been registered for treatment on the basis of results obtained using highly sensitive techniques of *KRAS* mutation testing in mCRCs [21, 103-106] and lung cancers [107]. Thus, high-sensitivity *KRAS* detection methods improve the prediction of benefit from targeted therapy, thereby justifying their use for routine *KRAS* testing.

Even in microdissected tumor material, mutant sequences in DNA isolate can be rare, and more sensitive methods than direct sequencing may be required for mutation analysis. One can imagine that a perfect approach would be the combination of manual microdissection (under the microscope) with high-sensitivity (1% mutants in a 99% wild-type background limit of detection) methods for *KRAS* mutation testing. For example, if a microdissected material contains approximately 90% cancer cells, a subpopulation of approximately 1-2% of mutant cells can be detected. The requirement of such a high level of sensitivity in detecting mutant tumor cells does not seem exceptional if we consider that, to predict the response to hormonal therapy in breast cancer, an ER/PR positive tumor cell level as low as 1% is required to expect a significant clinical response [108].

## Diagnostic assays for *KRAS* mutation testing, with the CE-IVD mark

After December 7, 2003, *in vitro* diagnostic (IVD) assays offered for sale in EU member countries are required to conform to IVD Directive requirements (98/79/EC) and to be CE-IVD-marked [109]. Tests without the CE-IVD mark are only acceptable for research. In the USA, similar but more restrictive FDA regulations have been in effect for many years. The CE-IVD mark guarantees that the product achieved the performance stated by the manufacturer on a variety of test parameters, such as limits of detection, analytical sensitivity and specificity, reproducibility/repeatability, and potential interfering and cross-reacting substances [110]. Nevertheless, it is surprising that, despite the EU regulations, direct sequencing and other "in-house" developed techniques were the most frequently used methods for clinical *KRAS* testing, as shown by the results of ESP-EQA 2011 program, in which the competence of participating laboratories (mainly from member states of the European Union) in providing clinically useful reports was assessed [111]. To the best of our knowledge, there are currently 12 commercially available CE-IVD-marked *KRAS* mutation kits available in Europe for diagnostic use that can be divided into three categories: real-time PCR-based, sequencing-based and DNA hybridization-based tests. The pros and cons of these assays are briefly described below.

***TheraScreen PCR and Pyro.*** The TheraScreen PCR (earlier DxS, now a Qiagen company) was the first clinically validated, CE-IVD-certified (2009) and FDA-approved (and currently the only commercially available *KRAS* test that is FDA-approved – for use on the Rotor-Gene Q instrument; 2012) diagnostic kit for the assessment of tumor-specific mutations in patients with CRC. It is a frequently used commercial *KRAS* test in Europe and the USA. In the ESP *KRAS* EQA scheme 2011, this kit was used by 23% of participants [111]. The kit detects six mutations in codon 12 and one mutation in codon 13 of the *KRAS* oncogene (Table I). The TheraScreen PCR kit (Qiagen, Manchester, UK) combines two methods, namely: ARMS (Amplification Refractory Mutation System) and Scorpions in real-time PCR reactions. In these PCR reactions, matched (mutant-specific) primers are efficiently amplified compared with mismatched primers (only low-level background amplification occurs). The real-time PCR assay is used to assess the quantity of mutant vs. wild-type sequences in the sample. An example of data acquisition is shown in Figure 1. The TheraScreen Pyro kit is the first clinically validated, CE-IVD-marked test for pyrosequencing-based detection of *KRAS* mutations in codons 12, 13 and 61 dedicated for use in the PyroMark Q24 system.

Recent reports suggest that the TheraScreen PCR kit has a significantly higher sensitivity compared with sequencing, even in samples with optimal tumor cell content [2, 61, 92-94]. Angulo *et al.* [61] compared the true sensitivity of direct sequencing with that of the TheraScreen PCR kit. When the mutant DNA represented 5% of the total DNA, the TheraScreen PCR kit detected mutations in 84% of the samples. This sensitivity was much higher than that obtained by direct sequencing, in which mutations were detected in 19 and 76% of the samples if mutant DNA constituted 5 and 30%, respectively. This test increases the detection percentage of *KRAS* mutations by 7-25% compared with direct sequencing [2, 92-94]. The limit of detection (LoD) of this assay (derived from cell lines) is 1% mutants in a 99% wild-type background [112]. Small intra- and inter-lot deviations and a good concordance among the different real-time PCR systems suggested the reliability of this test for clinical use [113]. Ad-

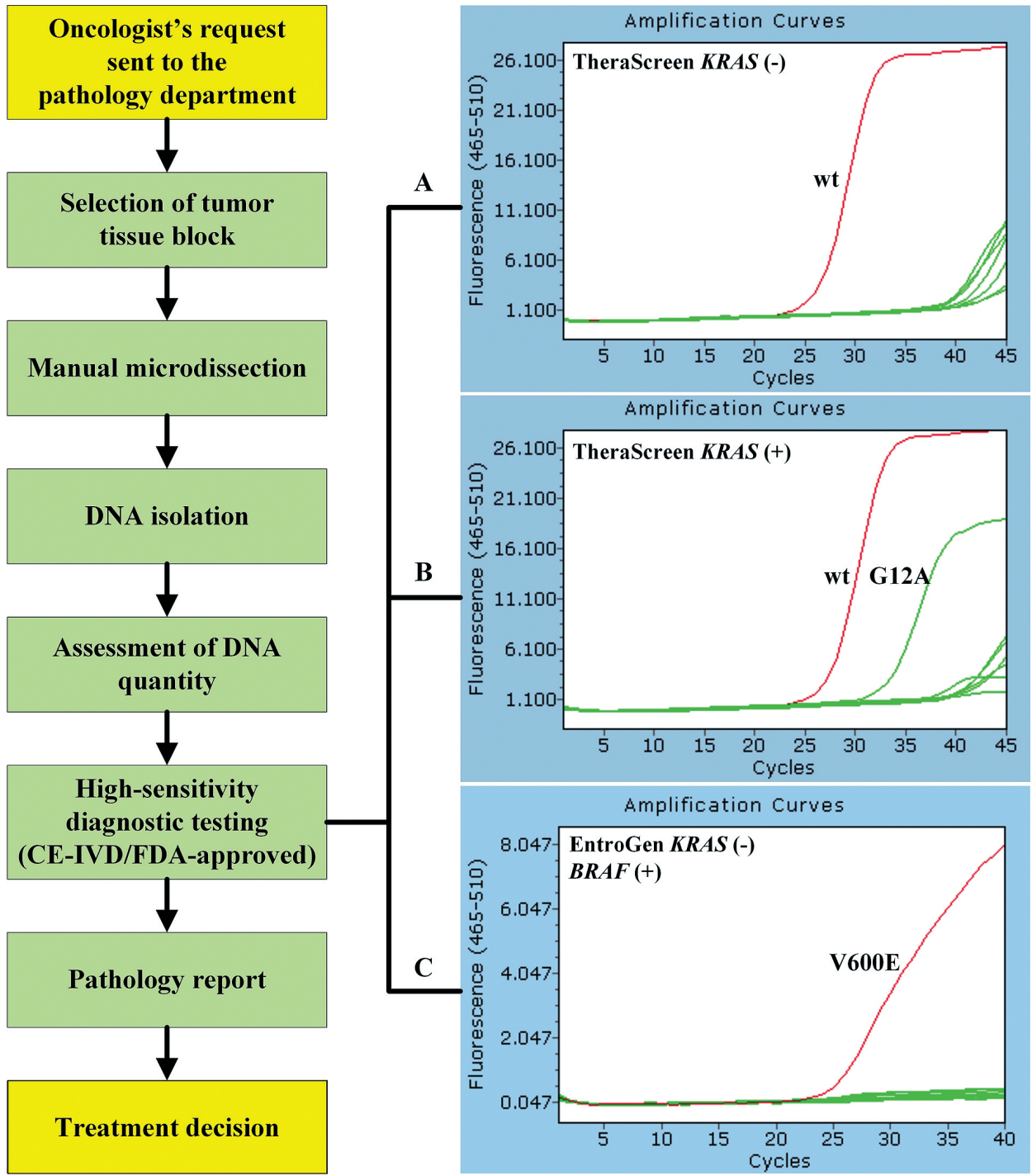

**Fig. 1.** Flow chart of the main stages in *KRAS* mutation testing for clinical use. Steps of the procedure performed by an oncologist (yellow background), and steps performed by a pathologist (green background). (A, B) An example of the raw amplification curves of *KRAS* wild-type (A) and mutant (B) samples tested using the TheraScreen PCR mutation kit. (A, B) The curve on the farthest left (red) represents the amplification product of the wild-type (wt) control DNA (exon 4). (B) The next curve to the right (green) represents the amplification product of the mutant template (p.G12A). The small difference between the Cp values of the wild-type control DNA and the mutant DNA (ΔCp = 4.7 cycles) indicates a mutation. (C) An example of the raw amplification curves (FAM fluorophore) of sample tested using the *KRAS*/*BRAF* EntroGen mutation kit. The red curve (Cp = 24.96) represents amplification product of the *BRAF* V600E mutant DNA (control DNA Cp = 23.68).

155

ditional advantages of the TheraScreen PCR kit include its fast turnaround time, user-friendly software-assisted objective data interpretation, and one-step system that prevents contamination [2].

The most important limitation of this kit is the relatively high cost of the test per sample (~143 USD [114]; ~210 USD in Poland; Table I) and high DNA input requirements (Fig. 2).

*PNAClamp.* The PNAClamp mutation detection kit (Panagene, Daejeon, Korea) is based on peptide nucleic acid analog (PNA)-mediated real-time PCR clamping technology. The higher specificity of PNA binding to DNA, higher stability of a PNA-DNA duplex compared with the corresponding DNA-DNA duplex, and inefficiency of PNA to act as a primer for DNA polymerases are the bases for this technique [115]. PNA will hybridize to its complementary DNA target sequence only if the sequence is a complete match. The PNA/DNA complex effectively blocks the formation of a PCR product. On the other hand, in the case of the mutant alleles, the melting temperature of mismatched PNA-DNA hybrid is much lower than the corresponding temperature of the normal PNA-DNA hybrid. Consequently, mutated sequences are preferentially amplified, and amplification of the wild-type gene is suppressed [115, 116].

PNA-mediated PCR clamping significantly increased the percentages of detected *KRAS*, *BRAF*, and *PIK3CA* mutations compared with direct sequencing. PNA-mediated PCR clamping increased the percentage of detected *KRAS* mutations by 7-11% [95, 96] which implies that this method can detect minor subpopulations of tumor cells with mutant alleles. The detection of *KRAS* mutations in tumors of low cellularity depends on the method used. Direct sequencing revealed *KRAS* mutations in 11 out of 114 lung cancers (10%), whereas 10 additional mutations were detected (18%) using PNA-mediated PCR clamping. Of these mutations, five were detected in samples with low tumor cellularity [97]. In addition, this method detected more occult metastases in lymph nodes than standard HE analysis [117]. In another study [98], the PNAClamp test detected mutations in 1% of the mutant cells, while direct sequencing barely found mutations in 20-50% of the mutant cells. The PNA-Clamp test is a very sensitive method for detecting mutants in a very small amount of DNA (optimal range: 10-25 ng of total DNA), and it is suitable for *KRAS* mutation testing in small biopsy specimens (Fig. 2) [88, 98, 118].

The most important limitation of this kit is its inability to determine type of mutation in codon 12.

*StripAssay.* A completely different type of test is the StripAssay (Vienna Labs, Vienna, Austria). This test combines mutant-enriched PCR based on PNA clamping and reverse-hybridization to nitrocellulose test strips containing specific probes for the different mutations, immobilized as an array of parallel lines. Bound biotinylated sequences are detected using streptavidin-alkaline phosphatase and color substrates with a LoD of 1% mutants in a 99% wild-type background [119]. Fifteen percent of the samples that were identified as *KRAS*-positive by this test were diagnosed as wild-type by direct sequencing [120]. This test only requires a water bath, fluorometer and thermocycler. Today, it is difficult to imagine the laboratory of molecular pathology that would not be equipped with these devices. The StripAssay test detects eight mutations in codon 12 and two mutations in codon 13 of the *KRAS* oncogene (Table I). It is difficult to explain why two very rare mutations (p.G12I and

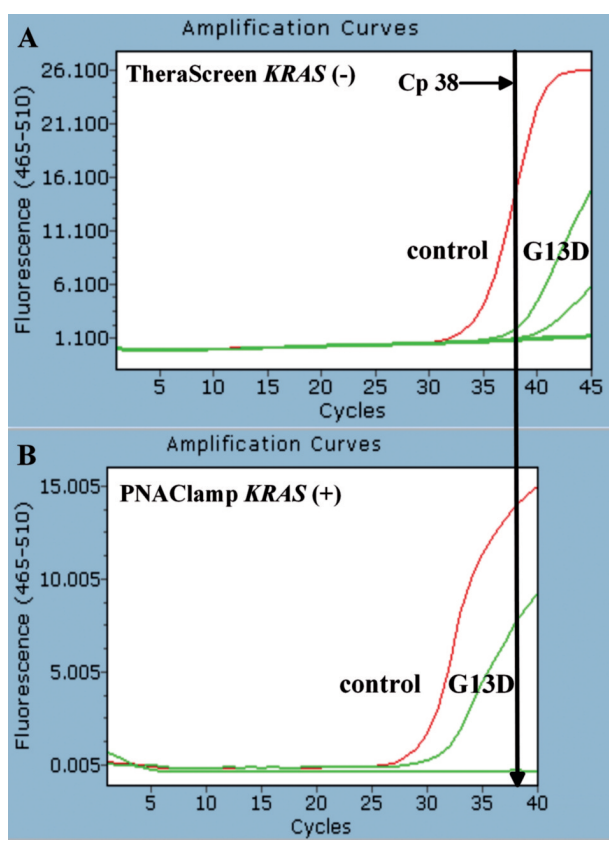

Fig. 2. An example of a discrepancy between the TheraScreen PCR kit and the PNAClamp kit for *KRAS* mutation testing resulting from insufficient DNA input for TheraScreen PCR kit. The same DNA input (27.3 ng total DNA) from the same patient was used for both kits. (A) The TheraScreen amplification curve for wild-type control DNA (Cp = 33.82) (red) indicates a low DNA input, thus very low-level mutations may not be detected (a warning is displayed in the test result). The next curve to the right (green) represents the amplification product for mutant DNA (p.G13D); nevertheless, if the Cp value is greater than or equal to 38 (as in this case), it is classified as negative or beyond the LoD of the kit and cannot be confidently interpreted as a mutation. (B) The same amount of DNA input is optimal for the PNAClamp kit (control Cp = 29.24; red) and a mutation in codon 13 (green) is unequivocally detected.

p.G12L) described in only five CRCs (according to the COSMIC database v60) are detected by this test, but other, much more frequent mutations are not.

The assay procedure is labor intensive (~6 h), because it requires several reaction steps: amplification, hybridization, detection, and washing. Nevertheless, the assay can be performed with standard laboratory equipment; therefore, this approach appears to be an interesting alternative to methods currently in use for the detection of *KRAS* mutations in DNA isolated from FFPE tissue [120].

*Cobas.* This is a TaqMelt-based real-time PCR assay designed to detect the presence of 21 *KRAS* mutations in codons 12, 13, and 61 (Table I) with a LoD of approximately 5% mutants in a 95% wild-type background [110]. Mutation detection is achieved by melting curve analysis, using the Cobas 4800 System (Roche, Branchburg, NJ, USA) with automated result interpretation software. This test is highly reproducible between different clinical laboratories (98% concordant results), requires 100 ng of total DNA input and is more sensitive than both sequencing and the TheraScreen test [121]. The additional 12 mutations detected by the Cobas test, compared with the TheraScreen kit, represent approximately 2% of all *KRAS* mutations. However, recent studies suggest that codon 61 mutations may be more prevalent than reflected in the COSMIC data [122, 123].

The most important limitation of this test is the need for dedicated equipment.

*Surveyor.* This technology is based on a mismatch-specific DNA endonuclease from celery (Surveyor nuclease), which cleaves with high specificity at the 3'side of any mismatch site in both DNA strands, including all base substitutions and insertion/deletions up to at least 12 nucleotides. Subsequently, DNA fragments are analyzed with the WAVE HS System [124]. The Surveyor kit (Transgenomic, Omaha, USA) detects all mutations in exon 2 of the *KRAS* gene (Table I) with a LoD of 1% mutants in a 99% wild-type background [125]. Sequencing is required to confirm base changes.

The most important limitation of this test is the need for dedicated equipment.

*LightMix.* This kit (TIB Molbiol, Berlin, Germany) uses real-time PCR clamping technology and the melting curve method, with a LoD of 1% mutants in a 99% wild-type background. Whitehall *et al.* [126] reported that this kit detected a higher number of mutations (59.5%) in FFPE tissue compared with fresh-frozen samples (38.8%). They also found incorrect *KRAS* mutation types in 27% of the samples. Whitehall *et al.* [126] concluded that although this kit performed reasonably well with frozen tissue DNA, it yielded an unacceptably high frequency of false-positive results with FFPE samples.

### Other (recently developed) assays

*EntroGen.* This kit (Tarzana, CA, USA) (Fig. 1.) is based on the ARMS-PCR method and detects 11 mutations (Table I) of the *KRAS* gene, with a LoD of 1% mutants in a 99% wild-type background. This test increased the detection percentage of *KRAS* mutations by 3% compared with direct sequencing [99].

*RealQuality RI-KRAS MuST.* This test (Ab Analitica, Padua, Italy) detects the 7 most frequent mutations (Table I) in the *KRAS* gene using real-time sequence-specific primer PCR assay. The LoD is 1% mutants in a 99% wild-type background [100]. This test increases the detection percentage of *KRAS* mutation by 9% compared with direct sequencing [100].

Currently there are no published scientific reports on the results of *KRAS* mutation testing in CRCs for the following three assays (kits). (1) *Hybcell Oncogenes Tissue plexA* (Anagnostics Bioanalysis, St. Valentin, Austria) is based on compact sequencing and principles of microarrays (called hybcell arrays). Processing is performed in a hyborg device. While traditional arrays are printed on slides, the hybcell arrays are composed of primers that are immobilized on the surface of a cylinder. The test detects 7 common mutations (Table I) with a LoD of 1-5% mutants in a 95-99% wild-type background. The extended version of the test includes an additional 25 *KRAS* mutations in codons 13, 61, and 146 and is under development. (2) *Devyser* (Devyser, Hägersten, Sweden) uses multiplex allele-specific PCR amplification for the identification of 12 *KRAS* mutations (Table I). The LoD is 3% mutants in a 97% wild-type background, according to the manufacturer's instructions. The PCR products are analyzed using a capillary electrophoresis genetic analyzer. (3) *AmoyDx* (Amoy Diagnostics, Xiamen, China) is based on the ARMS-PCR method and detects the 7 most common activating mutations (Table I) of the *KRAS* gene in cancer tissue. The LoD is 1% mutants in a 99% wild-type background, according to the manufacturer's instructions.

### Comparison of *KRAS* diagnostic assays

Data comparing different diagnostic assays (CE-IVD-marked) for *KRAS* mutation analysis are limited. A 95% concordance between the TheraScreen and StripAssay tests [127] as well as a 98% concordance between the TheraScreen and EntroGen tests [99] was reported. Bando *et al.*'s [2] study, which mimicked a real situation in a clinical genetic laboratory, demonstrated high overall agreement between the TheraScreen and StripAssay tests. However, they noted an excess of *KRAS* mutations found by the StripAssay test compared with the TheraScreen test, and they favored the interpretation that the StripAssay test exhibits a higher false-positive rate.

Paweł Domagała, Jolanta Hybiak, Violetta Sulżyc-Bielicka, *et al.*

In our view, the literature supports the statement that currently the **final choice of CE-IVD-marked *KRAS* mutation testing kit is largely dependent on the laboratory equipment, experience and test cost.** To compare the test cost per sample we obtained price quotes for kits available in Poland (August 2012). The results are shown in Table I.

Unfortunately, the cost per sample may depend on both the type of kit used and on the country where the kit is sold. For example, the cost per sample for the TheraScreen test is about USD 143 [114]. However, in Poland, the cost per sample of the same test is about USD 210 (kit cost obtained from Syngen Biotech, distributor of Qiagen in Poland). The high cost makes it difficult to use this test in routine clinical practice.

The number of commercially available diagnostic tests (CE-IVD-marked) will be increasing very rapidly. Real-time PCR-based methods are simple and sensitive, and will therefore be useful for molecular tests of predictive clinical significance performed in the molecular laboratories of pathology departments. Further studies by independent researchers are needed to compare these tests, identify their advantages and disadvantages and suggest recommendations.

## *KRAS* Quality Assurance Programs

In order to demonstrate competence and expertise in molecular testing (as with all other laboratory tests), quality assurance programs based on scientific evidence are needed. There are two basic types of responsibility that are different but complementary, namely responsibility for the clinical (pathological) report and responsibility for the technical aspects of the laboratory procedure. Both types of responsibility can be assessed by external quality assurance (EQA) programs. EQA programs should be preceded by the publication of the appropriate guidelines for a particular test by scientific professional societies. Such guidelines have already been published for *KRAS* mutation testing [46, 62].

**External quality assessment** is regarded as one of the essential steps in the validation of clinical tests [128]. Therefore, we believe that an evidence-based approach requires that laboratories performing *KRAS* tests (and other molecular predictive tests of clinical significance) be validated by EQA programs. Somatic mutation tests containing predictive information and thus of therapeutic consequences should be CE-IVD/FDA approved, as has already been done with several commercially available kits for testing *KRAS* mutations (see above). In this respect, the guidelines of the European Society of Pathology would aid in the selection of specific technologies for routine *KRAS* testing in different countries [59].

There is a need for EQA programs because:
1) Data comparing different *KRAS* mutation assays are limited.
2) Most but not all assays provide accurate results [129].
3) Currently, several commercially available assays are in vitro diagnostic-certified for the detection of somatic *KRAS* mutations in clinical practice and Conformité Européenne (CE)-marked. These assays can be easily used in molecular laboratories of pathology departments but their performance should be rigorously monitored.
4) An EQA program provides an opportunity for laboratories to evaluate their performance in *KRAS* testing and to compare it with that of other laboratories.
5) An EQA program may accelerate the harmonization and unification of *KRAS* mutation result reports to ensure that the reports are informative for oncologists and patients.
6) Participation in an EQA program may help to identify errors in genotyping and develop methods to eliminate them. The findings of the first phase of the development of a European EQA scheme for *KRAS* testing, in collaboration with the European Society of Pathology showed that only ten out of 13 laboratories labeled all 14 test cases correctly; the other laboratories made one to four mistakes. In addition, the laboratories involved in that test were selected based on their prominent position in the field and all but one frequently performed *KRAS* mutation testing for diagnostic purposes [47]. In another study, an evaluation of the concordance between *KRAS* assays performed by 6 different laboratories that tested 20 CRCs showed that consensus scores of 100% and ≥ 83% (in all or 5/6 laboratories) were reached in only 11/20 (55%) and 17/20 (85%) cases, respectively [130]. The main weakness of this study compared with other reports [129, 131] was that laboratories analyzed *KRAS* mutations from centrally isolated DNA but not from tumor sections on glass slides. Thus, the results did not mimic clinical testing procedures. In general, EQA programs validate the expertise of a pathology laboratory in performing particular molecular tests and also provide means for improvement. For example, although 10-15% of the laboratories made unacceptable mistakes during the first round of ESP *KRAS* EQA scheme, after feedback, the labs performed significantly better [58].
7) A list of departments/laboratories that successfully passed the ESP *KRAS* EQA scheme maintained on the ESP website and on websites of the National Societies of Pathology/Oncology (similar to the Italian AIOM and SIAPEC websites [62]) will help oncologists to choose the appropriate laboratory for *KRAS* mutation testing. Additionally, this information may help the appropriate institutions make decisions concerning reimbursement for the cost of the test.

Thus, we believe that to achieve universal, high-quality standards for *KRAS* mutation testing and reporting, i.e., to ensure consistency and accuracy of results, an evidence-based approach requires validation of the performance of *KRAS* mutation testing by various laboratories through EQA programs. In several European countries (Germany, France, Italy, Spain, Sweden, The Netherlands, and the United Kingdom), national *KRAS* EQA programs have been organized under the auspices and supervision of the National Associations of Pathology and Oncology [47, 59]. Four years ago, the European Society of Pathology, as an international independent body, initiated a European *KRAS* EQA scheme [46, 47, 63] in which laboratories that performed the tests could participate irrespective of the DNA extraction method and *KRAS* testing method used. The ESP *KRAS* EQA scheme evaluated the correct identification of *KRAS* mutations, percentage of tumor cells and reporting of results. The main purpose of this EQA scheme is to assess the quality of *KRAS* testing. An additional but equally important aim is to provide a means for improvement. The findings of the 2011 ESP *KRAS* EQA scheme showed that 124 laboratories from 29 countries participated and 72% reported all 10 genotypes correctly, including (for the first time) four laboratories from Poland [132]: 1. Maria Skłodowska-Curie Memorial Cancer Center and Institute of Oncology; 2. Jagiellonian University Medical College, Chair of Pathology; 3. Oncogene Diagnostics; 4. Pomeranian Medical University, Department of Pathology.

## Revealing targets for personalized targeted therapy is the pathologist's domain

We would like to stress that we share the views of Silke Lassmann and Martin Werner who recently stated that: "…the expertise in morphologically- and molecular-based analyses of human tissue specimens is the pathologist's treasure and will be of utmost clinical relevance in terms of personalized medicine" [133]. The development of personalized targeted therapy is difficult because of the complex network of interactions between multiple molecular pathways [134, 135]. Molecular alterations in these signaling pathways are just beginning to be understood in the context of therapeutic response. Pathologists are increasingly involved in the analyses of molecular signaling pathways to reveal targets of personalized targeted therapy, especially now in an era where mutation-specific antibodies useful in targeted therapy are *ante portas* (at the door) [136]. Personalized medicine means "The right drug, the right dose, for the right patient, at the right time" [137], i.e., the matching of a particular therapy to specific molecular features of a tumor. However, prior to the administration of targeted drugs to suitable patients, predictive molecular or immunohistochemical diagnostic tests must be conducted for both medical and economic reasons (because most targeted drugs are very expensive). As outlined above, the development of highly sensitive, simple, reliable, rapid, and compatible with FFPE tissue *KRAS* mutation assays, instead of adapting tissue sampling and processing to research-derived molecular biology protocols, proved to be clinically useful.

*KRAS* testing may be regarded as a testing ground indicating how pathologists of tomorrow should embrace effectively molecular technologies for the benefit of cancer patients. **As outlined above, pathologists have the knowledge and expertise to identify particular gene mutations using the appropriate available molecular tests. In addition, the ESP has developed an EQA scheme to validate the expertise of the laboratories that perform the molecular tests and confirm that they are competent in this field.** Indeed, this is an example of evidence-based diagnostics in action.

While the clinical effectiveness of some targeted therapies (including cetuximab for the first-line chemotherapy of mCRC) has been proven, the cost/benefit ratio of such treatments is still a matter of debate [138, 139]. There are currently no standards for the timing of *KRAS* mutation testing in CRC patients. Although performing *KRAS* testing on all CRCs at the time of surgery would be advisable (similarly to ER/PR HER2 testing in breast cancer), it may currently be difficult to defend such an approach from a cost/effective point of view. In France, the National Cancer Institute allocated €2.5 million for *KRAS* mutation screening only in mCRC patients, to the 2008 budget [140]. It seems reasonable to expect and suggest that the timing of *KRAS* mutation testing in CRC patients should depend on the likelihood of metastatic disease in a patient with CRC at the time of operation. Postoperative material from high-risk patients could be tested, due to increased likelihood of the need to establish their *KRAS* mutation status [31]. In any case, *KRAS* mutation testing should be performed at the time of diagnosis of metastatic CRC [25, 31].

## Conclusions

Pathologists should not be afraid of new molecular technologies because only those simplified to "kits" are currently certified for clinical purposes (and can easily be implemented). Sophisticated but very expensive molecular technologies (next generation sequencing and whole genome analysis) are currently only utilized in research. When those technologies are simplified and become much less expensive, they will also be accommodated by those pathologists who are familiar with the molecular tests now required in clinical oncology practice. Integrated histological diagnosis, en-riched by predictive information provided by molecular assessments, is nothing new to pathologists; what is new is the level of precision of predictive information provided by molecular analyses and the cost of molecular tests and targeted treatments. *KRAS* testing is just an example of a molecular test required for the implementation of targeted therapy in a particular patient population. Pathologists must be ready to perform the appropriate molecular and immunohistochemical tests to facilitate the treatment decisions made by oncologists for the benefit of patients.

In conclusion, based on the existing literature and our own experience we would like to make the following recommendations:

1. *KRAS* mutation testing (and other molecular tests of clinical predictive significance) should be performed by molecular laboratories in the pathology departments. This is because pathologists are responsible for the final pathology reports of CRCs, which should include predictive information. Sending paraffin blocks away for testing is not in the best interest of patients and is therefore not an optimal solution.
2. One of the commercially available CE-IVD-marked or FDA-approved "kits" should be used for *KRAS* mutation testing. The use of "in-house-developed methods" is not an optimal solution and should be avoided.
3. The sensitivity of direct sequencing for the analysis of *KRAS* mutations in FFPE tumor samples seems to be too low to enable its use as a routine clinical test. The use of direct sequencing is not an optimal solution and should be avoided because, with this method, up to approximately 25% of patients selected for anti-EGFR therapy will not benefit.
4. Laboratories in the European Union performing *KRAS* mutation testing for diagnostic (predictive) purposes should periodically and continuously participate in the ESP *KRAS* EQA scheme – a program organized by an independent international body. Participation should be obligatory. The previous experiences of existing national EQA programs should be exploited and utilized by organizers of the ESP *KRAS* EQA scheme to reach a consensus on this issue, which has very important consequences for patients, oncologists and pathologists.

## Conflict of interest

The authors declare that they have no conflict of interest.

**Address for correspondence:**

Prof. **Wenancjusz Domagała**, MD
Department of Pathology
Pomeranian Medical University
Unii Lubelskiej 1
71-252 Szczecin
phone/fax +48 91 487 00 32
e-mail: wenek@pum.edu.pl